\documentclass{article}
\begin{document}
\hoffset = -2 truecm \voffset = -2 truecm
\newcommand{\dfrac}[2]{\frac{\displaystyle #1}{\displaystyle #2}}
\title{Symmetry restoring phase transitions at high density in a 4D
Nambu-Jona-Lasinio model with a single order parameter \footnote{This project
supported by National Natural Science Foundation of China.} \\
}
\author{Bang-Rong Zhou\thanks{zhoubr@163bj.com}\\
Department of Physics, Graduate School of \\ the Chinese Academy of Sciences,
Beijing 100039, China \\
and \\
Chinese Center of Advanced Science and Technology \\ (World Laboratory), Beijing
100080, China}
\date{}
\maketitle
\begin{abstract}
High density phase transitions in a 4 dimensional Nambu-Jona-Lasinio  model
containing a single symmetry breaking order parameter coming from the
fermion-antifermion condensates are researched and expounded by means of both the
gap equation and the effective potential approach. The phase transitions are proven
to be second order at a high temperature $T$; however at $T=0$, they are first- or
second- order, depending on whether $\Lambda/m(0)$, the ratio of the momentum cutoff
$\Lambda$ in the fermion loop integrals to the dynamical fermion mass $m(0)$ at zero
temperature, is less than 3.387 or not. The former condition can not be satisfied in
some models. The discussions further show complete effectiveness of the critical
analysis based on the gap equation for second order phase transitions including
determination of the condition of their occurrence.
\end{abstract}
{\it PACS}: 11.10.Wx; 11.10.Lm; 11.15.Pg; 11.30.Er \\
{\it Keywords}: Nambu-Jona-Lasinio model, symmetry restoring at high density, gap
equation, effective potential, second and first order phase transition \\ \\
\section{Introduction}
\indent Nambu-Jona-Lasinio (NJL) model \cite{kn:1} is a good laboratory to research
symmetry restoring phase transition at finite temperature $T$ and finite chemical
potential $\mu$ \cite{kn:2}. It could simulate the phase transitions in Quantum
Chromodynamics (QCD) \cite{kn:3} and is directly related to some dynamical schemes
of electroweak symmetry breaking \cite{kn:4,kn:5}. It is known that in the NJL model
with a single order parameter which comes from fermion-antifermion $(f\bar{f})$
condensates, the phase transitions at a high critical temperature $T_c$ for fixed
$\mu$ are second order \cite{kn:6}. On the other hand, with regard to the phase
transitions in this class of models at a high chemical potential for a fixed $T$,
especially for $T\rightarrow 0$ case, less deep-going researches were made and the
conclusions failed to be made so simple and clear. Therefore, a more careful
investigation of this problem is certainly interesting. On the high density phase
transitions at $T=0$ in this class of NJL models, the authors of Ref. \cite{kn:7}
derived the effective potential from the imaginary-time formalism of thermal field
theory and briefly outlined the results. However, it seems that some more detailed
demonstrations are necessary for understanding these results and in addition, some
of the results there need to be clarified further. Actuated by the above motivates,
in this paper, we will make a thorough examination of the phase transitions at high
$\mu$ in this class of  $4D$ NJL model. The discussions will be made in the
real-time formalism of thermal field theory and involve both $T\neq 0$ and $T=0$
case. For second order phase transitions, we will make critical analysis of the
order parameter of symmetry breaking based on both the gap equation and the zero
temperature effective potential. For first order phase transitions we will
explicitly give the equations determining the critical curves. The condition which
distinguishes between second and first order phase transitions will be indicated
definitely. \\
\indent The Lagrangian of the model will be expressed by
$$
{\cal L}(x)=\sum_{k=1}^N\bar{\psi}^k(x)i\gamma^{\mu}\partial_{\mu}\psi_k(x)+
              \frac{g}{2}\sum_{k=1}^N[\bar{\psi}^k(x)\psi_k(x)]^2,
\eqno(1)$$ where $\psi_k(x)$  are the fermion fields with $N$ "color" components and
$g$ is the four-fermion coupling constant. It is indicated that the four-fermion
interactions in Eq.(1) could lead to only the fermion-antifermion condensates
$\langle\bar{\psi}\psi\rangle$. The conclusions reached from Eq.(1) are essentially
also applicable to more complicated models with a single order parameter coming from
the condensates $\langle\bar{\psi}\psi\rangle$. The discussions will be made in the
fermion bubble diagram approximation which is equivalent to the leading order of the
$1/N$ expansion. Since when $D=4$ the Lagrangian (1) is not renormalizable, we will
regard it only as a low energy effective field theory and examine its physical
results.  We will analyze second order high $\mu$ phase transitions based on the gap
equation in Sect.\ref{gap}; examine second and first order high $\mu$ phase
transitions by means of the zero temperature effective potential respectively In
Sects. \ref{effsecond} and
\ref{efffirst} and in Sect.\ref{conclusion} come to our conclusions.\\
\section{Second order high $\mu$ phase transitions -- gap equation
analysis}\label{gap} \indent Assume that dynamical fermion mass could be generated
by the fermion condensates induced by the four-fermion interactions in Eq.(1) at
$T=\mu=0$ and at some finite $T$ and $\mu$, then the corresponding mass term will
spontaneously break the discrete chiral symmetry $\chi_D$:
$\psi_k(x)\stackrel{\chi_D}{\longrightarrow}\gamma_5\psi_k(x)$ and the special
parities ${\cal P}_j$: $\psi_k(t,\cdots,x^j,\cdots) \stackrel{{\cal
P}_j}{\longrightarrow}\gamma^j\psi_k(t,\cdots,-x^j,\cdots) (j=1,2,3)$. After the gap
equation at $T=\mu=0$ is substituted, the gap equation at $T\neq 0$ can be
transformed to \cite{kn:6}
$$
\frac{1}{2}m^2(0)\ln\left[\frac{\Lambda^2}{m^2(0)}+1\right]=
\frac{1}{2}m^2\ln\left(\frac{\Lambda^2}{m^2}+1\right)+F_3(T,\mu, m), \eqno(2)$$
where $\Lambda$ is the four dimensional Euclidean momentum cutoff in the fermion
loop integrals, $m(0)$ and $m\equiv m(T,\mu)$ are respectively the dynamical fermion
mass at $T=\mu=0$ and at $T\neq 0$ and/or $\mu\neq 0$ and
$$
F_3(T,\mu,m)=4T^2[I_3(y,-r)+I_3(y,r)] \eqno(3)$$ with
$$
I_3(y,\mp r)=\frac{1}{2}\int_0^{\infty}\frac{dx}{\sqrt{x^2+y^2}}
\frac{x^2}{\exp(\sqrt{x^2+y^2}\mp r)+1}, \eqno(4)$$
 and $y=m/T$ and $r=\mu/T$. It is
easy to verify that $F_3(T,\mu,m)$ is an increasing function of $T$ and $\mu$ but a
decreasing function of $m$, and from Eq.(2), we can assume that at some critical
temperature $T_c$ and critical chemical potential $\mu_c$, the dynamical fermion
mass $m$ becomes zero, thus the symmetries $\chi_D$ and $\mathcal{P}_j \ (j=1,2,3)$
which are spontaneously broken at lower $T$ and $\mu$ will be restored. Setting $m=0
(y=0)$ in Eq.(2) we will obtain the critical equation
$$
 \frac{1}{2}m^2(0)\ln\left[\frac{\Lambda^2}{m^2(0)}+1\right]=
2T_c^2\int_0^{\infty}dx\left[ \frac{x}{e^{x-\frac{\mu_c}{T_c}}+1}+(-\mu_c\to
\mu_c)\right]. \eqno(5)$$ When taking the limit $m\rightarrow 0$ and obtaining
Eq.(5), we did not encounter any singularity, this means that $m$ may go to zero
continuously thus Eq.(5) will be the critical equation of a second order phase
transition. In fact, for a given high temperature $T$,  substituting Eq.(5) with
$T_c$ replaced by T into Eq.(2), then using the high temperature expansion of
$F_3(T,\mu,m)$\cite{kn:6}, we can obtain the critical behavior of $m^2$ near $\mu_c$
at high $T$
$$
m^2=(\mu_c^2-\mu^2)\sum_{p=1}^{\infty}\frac{\cosh (p\bar{r}_c)}{ p\bar{r}_c}/
\left[\ln\frac{\Lambda}{T\pi}-\frac{1}{2}+\gamma
               -h(T,\mu) \right],
\ {\mathrm{when}} \ \mu\sim \mu_c \ \ \mathrm{and} \ \ T \ \mathrm{is \; \ high},
\eqno(6)$$ with $\bar{r}_c=\mu_c/T$ and
$$
h(T,\mu)=7\zeta(3)(\frac{r}{2\pi})^2-31\zeta(5)(\frac{r}{2\pi})^4
+127\zeta(7)(\frac{r}{2\pi})^6, \eqno(7)$$ where $\zeta(s) \ (s=3,5,7)$ are the
Riemann zeta functions and $\gamma$ is the Euler constant.  Equation (6) indicates
that for high $T$, the phase transition at $\mu_c$ is second order.  For proving the
phase transitions at $\mu_c$ for low $T$, we first consider the limit case of
$T\rightarrow 0$.  In this case ($r\rightarrow \infty$) , the critical equation (5)
with $\mu_c\rightarrow \mu$ and $T_c\rightarrow T$ is reduced to
$$
\mu^2=\frac{1}{2} m^2(0)\ln\left[\frac{\Lambda^2}{m^2(0)}+1\right]\equiv\mu_{c0}^2,
\ \ {\rm when} \ T\rightarrow 0. \eqno(8)$$ Through changing the integral variable
by $z=(x^2+y^2)^{1/2}$ we may rewrite
$$ F_3(T,\mu,
m)=2m^2\int_1^{\infty}dz\left[\frac{\sqrt{z^2-1}}{e^{y(z-\alpha)}+1}+ (-\alpha\to
\alpha)\right], \eqno(9)$$ with $\alpha=\mu/m$ and obtain
$$
F_3(T=0,\mu,m)=\theta(\mu-m)\left[\mu\sqrt{\mu^2-m^2}-m^2\ln\frac{\mu+\sqrt{\mu^2-m^2}}{m}
\right]. \eqno(10)$$
By using Eqs.(8) and (10), we may reduce the $T\rightarrow 0$
limit of the gap equation (2) to
$$\begin{array}{rcl}
&m=m(0), &{\rm when} \ \mu\leq m(0) \\
&\mu_{c0}^2-\mu^2\sqrt{1-\dfrac{m^2}{\mu^2}}=
\dfrac{m^2}{2}\ln\dfrac{\Lambda^2+m^2}{(\mu+\sqrt{\mu^2-m^2})^2}, &{\rm when} \ \mu> m(0).\\
\end{array}
\eqno(11)$$ From Eq. (11) it can be proven that
$$\begin{array}{rcl}
\lim\limits_{T\to 0}\dfrac{\partial m}{\partial\mu} &=&
-2\sqrt{\mu^2-m^2}/m\left[\ln\dfrac{\Lambda^2+m^2}{(\mu+\sqrt{\mu^2-m^2})^2}
-\dfrac{\Lambda^2}{\Lambda^2+m^2}\right] \\
&=&-m\sqrt{\mu^2-m^2}/\left(\mu_{c0}^2-\mu\sqrt{\mu^2-m^2}-
\dfrac{m^2}{2}\dfrac{\Lambda^2}{\Lambda^2+m^2}\right) \\
 &\leq &0, \ {\rm when} \  m(0)\leq \mu < \mu_{c0}  \\
\end{array}
\eqno(12)$$ and
$$ \lim\limits_{T\to
0}\frac{\partial^2 m}{\partial\mu^2}<0, \ {\rm when} \  m(0) < \mu < \mu_{c0}.
\eqno(13)$$
Eq.(12) shows that in the $T\rightarrow 0$ limit, when $\mu=m(0)$,
$\partial m/\partial \mu =0$ and when $\mu=\mu_{c0}$ where $m=0$, $\partial
m/\partial \mu =-\infty$. From these results and Eq.(13), we can deduce that the
$m-\mu$ curve at $T=0$ is concave downward in the region $m(0)\leq\mu\leq\mu_{c0}$.
Near the critical point $\mu\sim \mu_{c0}$, we find out from Eq.(11) that
$$
m^2\simeq(\mu_{c0}^2-\mu^2)/\left[\ln(\Lambda/2\mu)-1/2\right], \ {\rm when} \ \
T\rightarrow 0. \eqno(14)$$
Eq.(14) indicates that in the $D=4$ NJL model with a
single order parameter, when $T=0$, the same as when $T$ is high, the symmetry
restoration phase transition at $\mu_c$ could be second order. This is different
from the high density phase transitions when $T=0$ in $D=2$ and $D=3$ GN model where
they are only first order
\cite{kn:8,kn:9,kn:10}. \\
\indent However, for consistence of Eq.(14) i.e. when $\mu<\mu_{c0}$, $m^2\geq 0$,
we must have $\ln(\Lambda/2\mu)-1/2 \stackrel{>}{\sim} \ln(\Lambda/2\mu_{c0})-1/2
\geq 0$ and this implies that
$$\mu^2_{c0}\leq \Lambda^2/4e
\eqno(15) $$ which is the condition in which a second order phase transition could occur
at $T=0$. \\
\section{second order phase transitions -- effective potential analysis}
\label{effsecond}
 \indent  The above conclusion that when the condition (15) is
satisfied the high density phase transition at $T=0$ is second order reached only by
the gap equation analysis can be verified by an effective potential analysis. It has
been proven \cite{kn:11} that the extreme value condition of the thermal effective
potential $V_{eff}^{(4)}(T,\mu,m)$ is
$$ \frac{\partial
V_{eff}^{(4)}(T,\mu,m)}{\partial m}=0 \eqno(16)$$
with
$$ \frac{\partial V_{eff}^{(4)}(T,\mu,m)}{\partial
m}=
\frac{m}{2\pi^2}\left\{\frac{m^2}{2}\ln\left(\frac{\Lambda^2}{m^2}+1\right)-\frac{m^2(0)}{2}
\ln\left[\frac{\Lambda^2}{m^2(0)}+1\right]+ F_3(T,\mu,m)\right\}. \eqno(17)$$ Hence
Eq.(16) just corresponds to the gap equation (2) multiplied by $m$. From Eqs. (17)
and (10) we can find out the effective potential $V_{eff}^{(4)}(T=0,\mu,m)$ at $T=0$
satisfying the condition $V_{eff}^{(4)}(T=0,\mu,m=0)=0$ expressed by
$$\begin{array}{rcl} V_{eff}^{(4)}(T=0,\mu,m)&=&\dfrac{1}{2\pi^2}\left\{
\dfrac{m^4}{8}\ln\left(\dfrac{\Lambda^2}{m^2}+1\right)
+\dfrac{m^2\Lambda^2}{8}-\dfrac{\Lambda^4}{8}\ln\left(1+\dfrac{m^2}{\Lambda^2}\right)
-\dfrac{m^2(0)m^2}{4}\ln\left[\dfrac{\Lambda^2}{m^2(0)}+1\right]
\right.\vspace{0.5cm}\\
&&\left.+\dfrac{\mu^4}{6}+\theta(\mu-m)\left[\dfrac{\mu
m^2}{4}\sqrt{\mu^2-m^2}-\dfrac{\mu}{6}(\mu^2-m^2)^{\frac{3}{2}}
-\dfrac{m^4}{4}\ln\dfrac{\mu +\sqrt{\mu^2-m^2}}{m}\right]\right\}.
\end{array}\eqno(18)$$
Eq.(18) is essentially identical to the effective potential in Ref. \cite{kn:7}
derived from the imaginary-time thermal field theory, except the extra
"normalization" condition $V_{eff}^{(4)}(T=0,\mu,m=0)=0$. From Eqs. (10) and (17) we
can directly obtain the extreme value condition of $V_{eff}^{(4)}(T=0,\mu,m)$
$$\begin{array}{rcl}
\dfrac{\partial V_{eff}^{(4)}(T=0,\mu,m)}{\partial m}&=&
\dfrac{m}{2\pi^2}\left\{\dfrac{m^2}{2}\ln\left(\dfrac{\Lambda^2}{m^2}+1\right)
-\dfrac{m^2(0)}{2}\ln\left[\dfrac{\Lambda^2}{m^2(0)}+1\right]+\theta(\mu-m)
\right. \vspace{0.5cm}\\
&&\left.\times\left[\mu\sqrt{\mu^2-m^2}-m^2\ln
\dfrac{\mu+\sqrt{\mu^2-m^2}}{m}\right]\right\}=0 \\
\end{array}\eqno(19)$$ and
$$\begin{array}{rcl}
\dfrac{\partial^2 V_{eff}^{(4)}(T=0,\mu,m)}{\partial m^2}&=&
\dfrac{1}{2\pi^2}\left\{\dfrac{3}{2}m^2\ln\left(\dfrac{\Lambda^2}{m^2}+1\right)
-m^2+\dfrac{m^4}{\Lambda^2+m^2}
-\dfrac{m^2(0)}{2}\ln\left[\dfrac{\Lambda^2}{m^2(0)}+1\right]\right.\vspace{0.5cm}\\
&&\left.+\theta(\mu-m)\left[\mu\sqrt{\mu^2-m^2}-3m^2\ln
\dfrac{\mu+\sqrt{\mu^2-m^2}}{m}\right]\right\}.
\end{array}\eqno(20)$$
Variation of $V_{eff}^{(4)}(T=0,\mu,m)$ as $\mu$ increases may be discussed by Eqs.
(18)-(20).\\
\indent First we note that when $\mu=0$, for any $\Lambda^2/m^2(0)\neq 0$,
$V_{eff}^{(4)}(T=0,\mu=0,m)$ will have a maximum point $m=0$ and a minimum point
$m=m(0)$ and this shows
spontaneous symmetry breaking at $T=\mu=0$. \\
\indent If the symmetries will be restored at high $\mu$ through second order phase
transitions, then the extreme value equation of $V_{eff}^{(4)}(T=0,\mu,m)$ with
$m\neq 0$, i.e. the $T\rightarrow 0$ limit of the gap equation (2), should have the
only solution and it must correspond to a minimum point of
$V_{eff}^{(4)}(T=0,\mu,m)$. We will examine this case in the condition expressed by
Eq.(15). This condition, by Eq.(8), is
$$\frac{m^2(0)}{2}\ln\left[\frac{\Lambda^2}{m^2(0)}+1\right]
\leq \frac{\Lambda^2}{4e}$$ which amounts to $\Lambda/m(0)\geq 3.387$ and will lead
to $\mu_{c0}>m(0)$. Variation of
$V_{eff}^{(4)}(T=0,\mu,m)$ as $\mu$ increases is as follows. \\
1) $0<\mu<m(0)$. In this case, $V_{eff}^{(4)}(T=0,\mu,m)$ will have a maximum point
$m=0({\rm when} \ m<\mu)$ and a minimum point $m=m(0)({\rm when} \ m>\mu)$. It can
be proven that the other extreme value equation contained in Eq. (19) when $m<\mu$
$$
\frac{1}{2}m^2(0)\ln\left[\frac{\Lambda^2}{m^2(0)}+1\right]=
\frac{m^2}{2}\ln\frac{\Lambda^2+m^2}{(\mu+\sqrt{\mu^2-m^2})^2}+\mu\sqrt{\mu^2-m^2}
\eqno(21)$$ has no solution for $m<\mu<m(0)$ when
$\frac{1}{2}\ln\left[\frac{\Lambda^2}{m^2(0)}+1\right]\geq 1$. In fact, if we set
$\mu=\alpha m(0)$ and $m=\beta\mu$, then Eq.(21) can be changed into
$$
\frac{1}{2}\ln\left[\frac{\Lambda^2}{m^2(0)}+1\right]=\frac{\beta^2\alpha^2}{2}\ln
\frac{\Lambda^2/m^2(0)+\alpha^2\beta^2}{\alpha^2(1+\sqrt{1-\beta^2})^2}
+\alpha^2\sqrt{1-\beta^2}. \eqno(22)$$ It is easy to check by numerical solution
that Eq.(22) has no solution for $\alpha<1$ and $\beta<1$ when
$\frac{1}{2}\ln\left[\frac{\Lambda^2}{m^2(0)}+1\right]\geq 1$ or
$\Lambda^2/m^2(0)\geq e^2-1$. Our prerequisite $\Lambda/m(0)\geq 3.387$ or
$\Lambda^2/m^2(0)\geq 15.11$ apparently satisfies this condition. Thus Eq.(21) does
have no solution. As a result, $m=m(0)$ becomes the only minimum point of
$V_{eff}^{(4)}(T=0,\mu,m)$ and we
will have the same spontaneous symmetry breaking as one at $T=\mu=0$. \\
2) $m(0)\leq \mu<\mu_{c0}$. In this case, when $m>\mu$, $V_{eff}^{(4)}(T=0,\mu,m)$
has no extreme value point and when $m<\mu$, $V_{eff}^{(4)}(T=0,\mu,m)$ will have a
maximum point $m=0$ and a minimum point $m_1$ which is determined by Eq. (21).
Noting that in present case Eq.(21) may have the solution $m=m_1$ when
$\frac{1}{2}\ln\left[\frac{\Lambda^2}{m^2(0)}+1\right]\geq 1$ and
$$\left.\frac{\partial^2 V_{eff}^{(4)}(T=0,\mu,m)}{\partial
m^2}\right|_{m=m_1}=\frac{1}{\pi^2}\left(\mu_{c0}^2-\mu\sqrt{\mu^2-m_1^2}
-\frac{m_1^2}{2}\frac{\Lambda^2}{\Lambda^2+m_1^2}\right)>0, \ \  {\rm when} \
\mu<\mu_{c0}.$$
In fact, Eq.(21) has the solutions $$ m_1=m(0), {\rm when} \
\mu=m(0) \eqno(23)$$ and
$$ m_1<m(0), {\rm when} \ \mu>m(0),
\eqno(24)$$ noting that equation (21) obeyed by $m_1$ is just the second formula in
Eq.(11), hence we can obtain from Eq.(12) that $\partial m_1/\partial \mu \leq 0$.
Equation (23) indicates that when $\mu=m(0)$, we will go back to the same case as
$\mu<m(0)$ and Eqs. (21) and (24) show that as $\mu$ increases from $m(0)$ further,
the minimum point $m_1$ will become smaller and smaller and finally it will go to
zero continuously at a critical chemical potential. Correspondingly, the global
minimum of $V_{eff}^{(4)}(T=0,\mu,m)$ at $m=m_1$ will go up as $\mu$ increases from
a negative value to zero, since it can be proven that
$$\frac {d V_{eff}^{(4)}(T=0,\mu,m_1)}{d\mu}=\frac{1}{12\pi^2}\left[
3m_1^2\sqrt{\mu^2-m_1^2}+4\mu^3-(\mu^2-m_1^2)^{3/2}
-3\mu^2\sqrt{\mu^2-m_1^2}\right]\geq 0.$$ The critical chemical potential can be
determined by taking $m=0$ in Eq. (21) and the result is precisely the $\mu_{c0}$
given by Eq.(8). Since Eq.(21) is exactly the second formula in Eq.(11) and the
critical behavior (14) of $m^2$ follows. Equation (8) expresses a critical curve
$C_2$
of second order phase transitions in $\mu-m(0)$ plane. \\
3) $\mu=\mu_{c0}$. Now $V_{eff}^{(4)}(T=0,\mu,m)$ has the only extreme value point
$m=0$ and it is found out that
$$\left.\frac{\partial^n V_{eff}^{(4)}(T=0,\mu,m)}{\partial
m^n}\right|_{m=0}=\left\{ \matrix{
 0, & {\rm when} \ n=2,3,5 \cr
 \dfrac{3}{2\pi^2}\left(\ln\dfrac{\Lambda^2}{4\mu_{c0}^2}-1\right),
               & {\rm when} \ n=4 \cr
 \dfrac{3}{2\pi^2}\left(\dfrac{4}{\Lambda^2}+\dfrac{5}{\mu_{c0}^2}\right),
               & {\rm when} \ n=6 \cr}\right..
\eqno(25) $$ Equation (25) implies that in the condition
$\ln(\Lambda^2/4\mu_{c0}^2)-1\geq 0$ or $\mu_{c0}^2\leq \Lambda^2/4e$, $m=0$ will be
the only minimum point of $V_{eff}^{(4)}(T=0,\mu,m)$
and the broken symmetries will be restored. \\
4) $\mu>\mu_{c0}$. In present case, it can be proven that the extreme value equation
(21) has no solution when $m<\mu$.  This can be seen most simply from Eq.(14) which
is the approximation of Eq.(21) when $\mu\sim \mu_{c0}$ and $m\approx 0$. More
rigorously, we can set $\mu=\gamma \mu_{c0}$, $m=\beta \mu=\beta\gamma\mu_{c0}$,
then Eq.(21) can be changed into
$$\frac{1}{2}\beta^2\gamma^2\left\{\ln\left(\frac{\Lambda^2}{\mu_{c0}^2}
+\beta^2\gamma^2\right)-2\ln[\gamma(1+\sqrt{1-\beta^2})]\right\}+\gamma^2\sqrt{1-\beta^2}=1
\eqno(26)$$ It may be checked that Eq.(26) has no solution with $\gamma>1 \
(\mu>\mu_{c0})$ and $\beta<1 \ (m<\mu)$ when $\mu_{c0}^2\leq \Lambda^2/4e$.
Consequently, $V_{eff}^{(4)}(T=0,\mu,m)$ will have the only minimum point $m=0$ in
this case and this fact further indicates restoration of the symmetries which were
broken at $T=0$ and $\mu<\mu_{c0}$ through a second order phase transition when
Eq.(15) is satisfied. \\
\indent Up to now we have proven that the phase transitions at $\mu_c$ are second
order for a high $T$ and for $T=0$ when $\mu^2_{c0}\leq \Lambda^2/4e$. Considering
these results and the fact that Eq.(5) is a critical equation of second order phase
transitions for any finite $T_c$ and $\mu_c$, we can conclude that the phase
transitions at $\mu_c$ for any $T$ including low $T$ are second order when
$\mu^2_{c0}\leq \Lambda^2/4e$. \\
\section{First order phase transition at $T=0$}\label{efffirst}
\indent A first order phase transition could generally occur in the case when the
extreme value equation of $V_{eff}^{(4)}(T=0,\mu,m)$ has two or more solutions with
$m\neq 0$. We will prove that is the case when $\mu_{c0}^2>\Lambda^2/4e$ and
$\mu>\mu_{c0}$. First we indicate that when $\mu>\mu_{c0}$, $m=0$ is always a
minimum point of $V_{eff}^{(4)}(T=0,\mu,m)$ and when $\mu_{c0}^2>\Lambda^2/4e$ ,
Eq.(26) will have the solutions with $\gamma>1$ and $\beta<1$, i.e. the extreme
value equation (21) will always have solutions for $\mu>\mu_{c0}>\Lambda/2e^{1/2}$
and $m<\mu$. This conclusion can also be checked by examining Eq.(14). Since in this
case $m=0$ is a minimum point, the solutions of Eq.(21) must correspond to at least
a maximum and a minimum point of $V_{eff}^{(4)}(T=0,\mu,m)$ and this could lead to a
first order phase transition. Assume the maximum point is $m_2$ and the minimum
point is $m_1$, then a first order phase transition curve should be determined by
the equations
$$V_{eff}^{(4)}(T=0,\mu,m=m_1)=V_{eff}^{(4)}(T=0,\mu,m=0)=0,
\eqno(27)$$
$$\left.\frac{\partial V_{eff}^{(4)}(T=0,\mu,m)}{\partial
m}\right|_{m=m_1\neq 0}=0 \eqno(28)$$ and
$$\left.\frac{\partial^2 V_{eff}^{(4)}(T=0,\mu,m)}{\partial
m^2}\right|_{m=0}>0 \eqno(29)$$ whose explicit or equivalent forms are
$$m_1^4\ln\left(\frac{\Lambda^2}{m_1^2}+1\right)+\Lambda^2m_1^2
-\Lambda^4\ln\left(1+\frac{m_1^2}{\Lambda^2}\right) -4m_1^2\mu_c^2+\frac{4}{3}\mu^4
+\theta(\mu-m_1)  $$ $$ \times\left[ 2\mu
m_1^2\sqrt{\mu^2-m_1^2}-\frac{4}{3}\mu(\mu^2-m_1^2)^{3/2}
-m_1^4\ln\frac{(\mu+\sqrt{\mu^2-m_1^2})^2}{m_1^2}\right]=0, \eqno(30)$$
$$\mu_{c0}^2=\frac{m_1^2}{2}\ln\left(\frac{\Lambda^2}{m_1^2}+1\right)
+\theta(\mu-m_1)\left[
\mu\sqrt{\mu^2-m_1^2}-m_1^2\ln\frac{\mu+\sqrt{\mu^2-m_1^2}}{m_1} \right],
\eqno(31)$$ where $m_1\neq 0$  and $\mu>\mu_{c0}$.  When $\mu\leq m_1$, from
Eqs.(31) and (8) we obtain $m_1=m(0)$. It may be pointed out that the condition
$\mu_{c0}^2>\Lambda^2/4e$ means that $\Lambda/m(0)<3.387$ thus we can have either
$\mu_{c0}>m(0)$ or $\mu_{c0}<m(0)$, so even if $\mu>\mu_{c0}$, it is still possible
for $\mu\leq m(0)$. Substituting $\mu\leq m_1=m(0)$ into Eq.(30), we are led to that
$$\frac{4}{3}\mu^4=m^4(0)\left[\ln(a+1)-a+a^2\ln\left(1+\frac{1}{a}\right)\right],
\ a\equiv \frac{\Lambda^2}{m^2(0)}, \eqno(32)$$ which expresses a first order phase
transition curve $C_1$ in the $\mu-m(0)$ plane. Eq.(32) is valid only if $\mu\leq
m_1=m(0)$, this will lead to the constraint
$\ln(a+1)-a+a^2\ln\left(1+\frac{1}{a}\right)\leq 4/3$ thus $\Lambda/m(0)\leq 2.21$.
So the curve $C_1$ must be in the region of $m(0)\geq \Lambda/2.21$.\\
\indent When $\mu>m(0)$, Eqs. (30)-(31) can be reduced to
$$\frac{4}{3}\mu^4-\frac{4}{3}\mu(\mu^2-m_1^2)^{3/2}=
2m_1^2\mu_{c0}^2-\Lambda^2m_1^2-\Lambda^4\ln\left(1+\frac{m_1^2}{\Lambda^2}\right),
\eqno(33)$$
$$
\mu_{c0}^2= \frac{m_1^2}{2}\ln\frac{\Lambda^2+m_1^2}{(\mu+\sqrt{\mu^2-m_1^2})^2}
+\mu\sqrt{\mu^2-m_1^2}.  \eqno(34)$$ Equations (33) and (34) together with $m_1\neq
0$ and $\mu> \mu_{c0}$ will be the equations to determine the first order phase
transition curve $C'_1$. When $\mu\rightarrow m(0)$, we may obtain from Eq.(34) that
$m_1\rightarrow m(0)$, as a result, equation (33) becomes
$$\left.\frac{4}{3}\mu^4\right|_{\mu\rightarrow m(0)}
=m^4(0)\left[\ln(a+1)-a+a^2\ln\left(1+\frac{1}{a}\right)\right],$$ which coincides
with Eq.(32) of the curve $C_1$ at the point $\mu=m(0)$. This means that the two
first order phase transition curves $C'_1$ and $C_1$ meet at the point $\mu=m(0)$.
On the other hand, we note that $\mu\rightarrow \mu_{c0}$ from $\mu>\mu_{c0}$ and
$m_1\rightarrow 0$ is a limiting solution of Eqs.(33) and (34) and that the curve
$C'_1$ should be in the region $\mu_{c0}^2>\Lambda^2/4e$, so it must intersect the
second order phase transition curve $C_2$ at $\mu_{c0}^2=\Lambda^2/4e$.
Consequently, in the $\mu-m(0)$ plane the curve $C'_1$ will start from the point
$\mu_{c0}^2=\Lambda^2/4e$, extend itself in the region of $\mu>\mu_{c0}$ and end at
$\mu=m(0)$ where it meets the curve $C_1$. The curve $C'_1$ must be in the region of
$\mu>\mu_{c0}$, since for the opposite case of $\mu\leq \mu_{c0}$, it may be seen
from Eqs. (20)and (25) that when $\mu_{c0}^2>\Lambda^2/4e$, $m=0$ will be a maximum
point of $V_{eff}^{(4)}(T=0, \mu, m)$, and this only corresponds to the phase of
symmetries being broken. The above discussions show that as far as symmetry
restoring phase transition is concerned, $\mu^2=\mu_{c0}^2=\Lambda^2/4e$ will be a
tricritical point in the $\mu-m(0)$ plane. We have noted that the similar results
were obtained in Ref. \cite{kn:7}, however, no detained demonstration leading to the
results, especially no distinction and relation between the first order phase
transition curves $C_1$ and $C'_1$ were given there. \\
\section{Conclusions}\label{conclusion}
\indent In this paper, we have researched the phase transitions at high density in a
4D NJL model with a single order parameter of symmetry breaking coming from
fermion-antifermion condensates by means of the gap equation and the zero
temperature effective potential. We have proven that for high $T$ the symmetry
restoring phase transitions are always second order; and for $T=0$, depending on
whether $\mu_{c0}^2$ is bigger than $\Lambda^2/4e$ or not, or equivalently, whether
$\Lambda/m(0)$ is less than $3.387$ or not, the phase transitions will be first- or
second- order. Hence $\mu_{c0}^2=\Lambda^2/4e$ is a tricritical point of high
density phase transitions at $T=0$. We have also further deduced that the phase
transitions at $\mu_c$ for any $T$ are second order when
$\Lambda/m(0)\geq 3.387$. \\
\indent We note that in this class of models, first order phase transitions occur
only if the dynamical fermion mass $m(0)$ at $T=0$ is close to and has the same
order of magnitude as the momentum cutoff $\Lambda$ of the loop integrals. However,
this condition is generally not natural or can not be satisfied in some low energy
effective theories of NJL-form. For instance, in the top-quark condensate scheme of
electroweak symmetry breaking \cite{kn:4}, $\Lambda/m(0)$ must be so large that up
to $\sim 10^{11}$ owing to the constraint from the standard electroweak model. Even
if in the four-generation fermion extension of such scheme \cite{kn:5} where
$\Lambda/m(0)$ could go down greatly, it must still be bigger than 10. Therefore, in
some physical 4D NJL models, it can be assumed that first order phase transition in
fact does not occur and even if at $T=0$, there are only second order phase transitions. \\
\indent It is seen from the above discussions that as far as analysis of a second
order phase transition is concerned, the gap equation approach and the effective
potential approach have the same effectiveness, including determination of the
occurrence condition of a second order phase transition.  However, the critical
analysis of the order parameter $m$ based on the gap equation is apparently more
simple and direct than the ones based on the effective potential.  \\
\indent The total conclusions in this paper are obtained only in the model (1) with
a single order parameter coming from the fermion-antifermion condensates
$\langle\bar{\psi}\psi\rangle$ and they could have substantial change when one
transfers to the model with both the fermion-antifermion condensates
$\langle\bar{\psi}\psi\rangle$ and the fermion-fermion condensates
$\langle\psi\psi\rangle$. The latter model may be a better simulation to QCD and
deserves to be researched further. \\
\indent The author is grateful to Professor
K.G. Klimenko for a private communication which brought the author's attention to
the relevant work in Ref.\cite{kn:7}.


\begin{thebibliography}{99}
\bibitem{kn:1}  Y. Nambu and G. Jona-Lasinio, Phys. Rev. {\bf 122}(1961)345;
                {\bf 124} (1961)246.
\bibitem{kn:2}  D. A. Kirzhnits and A. D. Linde, Phys. Lett.  {\bf 42B}
                (1972)471;
                S. Weinberg, Phys. Rev. D {\bf 7} (1973) 2887;
                {\bf 9} (1974) 3357;
                L. Dolan and R. Jackiw, {\it ibid.} {\bf 9} (1974) 3320;
                A. D. Linde, Rep. Prog. Phys. {\bf 42} (1979) 389;
                R. H. Brandenberger, Rev. Mod. Phys. {\bf 57} (1985) 1;
\bibitem{kn:3}  S. P. Klevansky, Rev. Mod. Phys.
                {\bf 64} (1992) 649; T. Hatsuda and T. Kunihiro, Phys. Rep.
                \textbf{247} (1994) 221; M.Alford, K.Rajagopal and F. Wilczek,
                Phys. Lett. B{\bf 422} (1998) 247; R. Rapp,T.Schafer, E.V.
                Shuryak and M. Velkovsky, Phys. Rev. Lett. {\bf 81} (1998) 53;
\bibitem{kn:4} Y. Nambu, in {\it New Theory in Physics, Proceedings of XI
               International Symposium on Elementary Particle Physics},
               Kazimierz, Poland, 1988, eds Z. Ajduk, S. Pokorski and
               A. Trautman, World Scientific, Singapore (1989);
               V.A. Miransky, M. Tanabashi and K. Yamawaki, Mod. Phys. Lett.
               {\bf A4} (1989) 1043; Phys. Lett. {\bf B221} (1989) 177.
               W.A. Bardeen, C.T. Hill and M. Lindner, Phys. Rev. {\bf D41}
               (1990) 1647.
\bibitem{kn:5} B.R. Zhou, Phys. Rev. D{\bf 47} (1993) 2656; D{\bf 47} (1993) 5038.                J. Berges and K. Rajagopal, Nucl Phys. B{\bf 538} (1999) 215.
\bibitem{kn:6} For example, see B. R. Zhou, Phys. Rev. D {\bf 57} (1998) 3171;
                Commun. Theor. Phys. {\bf 32} (1999) 425.
\bibitem{kn:7} D. Ebert, K. G. Klimenko, M.A. Vdovichenko and A.S.
               Vshivtsev, Phys. Rev. D{\bf 61} (2000) 025005.
\bibitem{kn:8}  U. Wolff, Phys. Lett. B{\bf 157} (1985) 303;
                F. Karsch, J. Kogut and H. W. Wyld, Nucl. Phys.
                B{\bf 280} (1987) 289;
                K. G. Klimenko, Z. Phys. C{\bf 37} (1988) 457;
                A. Chodos and H. Minakata, Phys. Lett. A{\bf 191} (1994) 39;
                Nucl. Phys. B{\bf 490} (1997) 687;T. Inagaki, T. Kouno and T.
                Muta, Inter. J. Mod. Phys. A{\bf 10} (1995) 2241;
                A. Chodos, F. Cooper, W. Mao, H. Minakata and A. Singh, Phys.
                Rev. D{\bf 61} (2000) 045011.
\bibitem{kn:9}  S. Hands, S. Kim and J. B. Kogut, Nucl. Phys. B{\bf 442} (1995) 364;
                S. Hands, Nucl. Phys. A{\bf 642} (1998) 228c;
                J. B. Kogut and C. G. Strouthos, Phys. Rev. D{\bf 63} (2001) 054502.
\bibitem{kn:10} B.R. Zhou, Commun. Theor. Phys. {\bf 40} (2003) 67.
\bibitem{kn:11} B.R. Zhou, Commun. Theor. Phys. {\bf 39} (2003) 663.
\end{thebibliography}
\end{document}